\documentclass{aa}  
\usepackage{graphicx}
\usepackage[varg]{txfonts}
\usepackage{longtable,lscape}
\begin{document}
   \title{Galaxy distribution and evolution around \\a sample of 2dF groups}


   \author{A.L.B. Ribeiro
          \inst{1}, M. Trevisan\inst{2}, P.A.A. Lopes\inst{3} \and A.C. Schilling\inst{1}}
        

   \offprints{A.L.B. Ribeiro}

   \institute{Laborat\'orio de Astrof\'{\i}sica Te\'orica e Observacional\\ 
              Departamento de Ci\^encias Exatas
e Tecnol\'ogicas\\ Universidade Estadual de Santa Cruz -- 45650-000, Ilh\'eus-BA, Brazil\\
              \email{albr@uesc.br}\\
		\email{ana.schilling@gmail.com}
              \and Instituto Astron\^omico e Geof\'{\i}sico- USP, S\~ao Paulo-SP, Brazil\\
\email{trevisan@iagusp.usp.br}
\and IP\&D Universidade do Vale do Para\'{\i}ba, Av. Shishima Hifumi 2911, \\
S\~ao Jos\'e dos Campos, SP 12244-000, Brasil \\
\email{paal05@gmail.com}}

   \date{Received April 29, 2009; Accepted XXXX, 2009}

 
  \abstract
   {We study galaxy evolution and spatial patterns in the surroundings of a sample of 2dF groups.}
   {Our aim is to find evidence of galaxy evolution and clustering out to 10 times
the virial radius of the groups and so redefine their properties
according to the spatial patterns in the fields and relate them to galaxy evolution.}
   {Group members and interlopers were redefined after the
identification of gaps in the redshift distribution. We then used exploratory spatial statistics based on the
the second moment of the Ripley function to probe the anisotropy in the galaxy distribution around the groups.}
   {We found an important anticorrelation between anisotropy around groups and the 
fraction of early-type galaxies in these fields. Our results illustrate how the dynamical state of galaxy
groups can be ascertained by the systematic study of their neighborhoods.
This is an important achievement, since the correct estimate of the extent to which
galaxies are affected by the group environment and follow large-scale filamentary structure is
relevant to understanding the process of galaxy clustering and evolution in the Universe.}
   {} 
   \maketitle
%

\section{Introduction} 

Small groups of galaxies contain about half of all galaxies in the
Universe (e.g., Huchra \& Geller 1982; Geller \&
Huchra 1983; Nolthenius \& White 1987; Ramella et al. 1989). 
They represent the link between galaxies and large-scale structures, 
and have at least two
important features: galaxies inside groups interact more
with each other than they do in the field; and groups
have small crossing times, generally $\lesssim 0.1~H_0^{-1}$,
indicating that they are dynamical units, that are possibly in
virial equilibrium. However, the dynamical state of a galaxy group
is not easy to determine. Group environments are unstable, the
systems still may be separating from the cosmic expansion, collapsing,
accreting new members, or merging with other groups to produce larger objects.
Generally, the estimated properties of these
systems are based on the assumption that groups of galaxies
defined by friends-of-friends algorithm (and other
clustering methods) are gravitationally
bound objects. This is not completely true, since projection effects
can dominate the statistics of these systems (e.g., Diaferio et al. 1993).
In a ${\rm\Lambda  CDM}$ cosmology, Niemi et al. (2007) showed that about
20\% of nearby groups are not bound, but merely visual objects.
There is no methodology for determining the dynamical status
of galaxy groups. An interesting attempt
at describing group evolution is the fundamental
track diagram, a plane that follows the evolution of isolated galaxy systems
in an expanding Universe (see Mamon 1993). The plane is defined by the
dimensionless crossing time and the dimensionless mass bias.
This diagram, however, suffers
from degeneracies between the expansion and early collapse phases,
and also between the full collapse and rebound phases.
Although most groups lie close to the fundamental track,
there is a large scatter and the result is inconclusive (Mamon 2007).
Giuricin et al. (1988) applied a correction
factor to the virial mass and assumed a specific model for the 
system evolution, but the model only accounted for internal
gravitational forces and neglected tidal interactions with 
the neighborhoods. Galaxy groups, however, 
interact significantly with their surroundings. For instance, the shapes and galaxy
flows around these systems are related to large-scale structures
and are relevant to the internal dynamics of the groups
(e.g., Paz et al. 2006; Ceccarelli et al. 2005; Plionis et al. 2004).
Generally, one assumes that bound groups reach a quasi-equilibrium 
state in which galaxies have isotropic orbits with random phases. 
This happens after the scattering of galaxies by each other and by 
masses outside the group. If a group is isolated
and remains fairly spherical, then its constituent galaxies are not 
deflected from their radial trajectories until the group has 
collapsed to a small fraction of its maximum radius. 
In this case, the collapse is violent and the group first reaches
equilibrium at $\sim$200 times the mean cosmic density. 
At the opposite extreme in which a nascent group is strongly 
influenced by surrounding objects, the colapse
is gentle and the group attains equilibrium at lower density constrasts.
Hence, measuring environmental influence over galaxy systems can be
a way of accessing their formation history and present dynamical state.
In the present work, this important point is investigated
where we study the surroundings of galaxy groups previously
selected from 2dF by Tago et al. (2006). Using some tools of spatial
statistical analysis, we examine the possible correlation between the anisotropy
around groups and galaxy evolution. This relationship may shed some light
on the dynamical state of galaxy groups.

This paper is organized as follows: in Sect. 2 we present the methodology and
data used in this work; in Sect. 3, we present our results and explore the relation between anisotropy and galaxy evolution; in Sect. 4, we discuss our results.
 
\section{Methodology and data}
\subsection{Probing anisotropy around groups}

For the projected distribution, 
anisotropy can be probed by the reduced second-order moment measure $\mathcal{K}$
of a point pattern (e.g., Stoyan et al. 1995).
In this work, we estimate $\mathcal{K}$ using the library {\bf spatstat} (see Baddeley 2008)
within the R statistical package.
The command Kmeasure (spatstat) executes the following 
steps:

\begin{enumerate}

\item A point pattern is assumed.
\item A list of all pairs of distinct points in the pattern is produced.
\item The vectors that join
the first point to the second point in each pair are computed.
\item These vectors are considered to be a new pattern of `points'. 
\item A Gaussian kernel smoother is applied to them.

\end{enumerate}

\begin{figure}
\centering
\includegraphics[width=8.5cm,height=8.5cm]{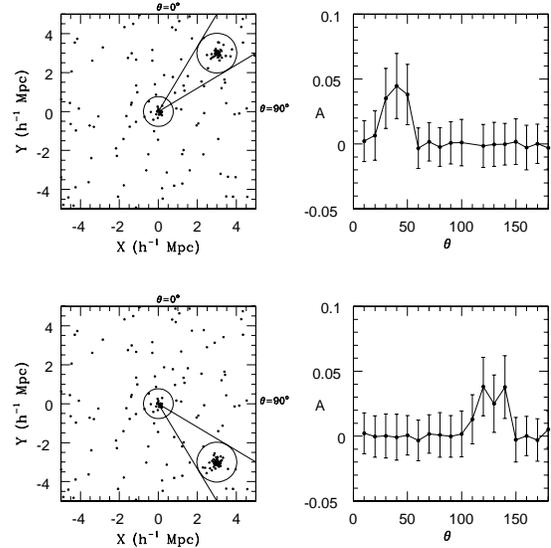}
\includegraphics[width=7cm,height=7cm]{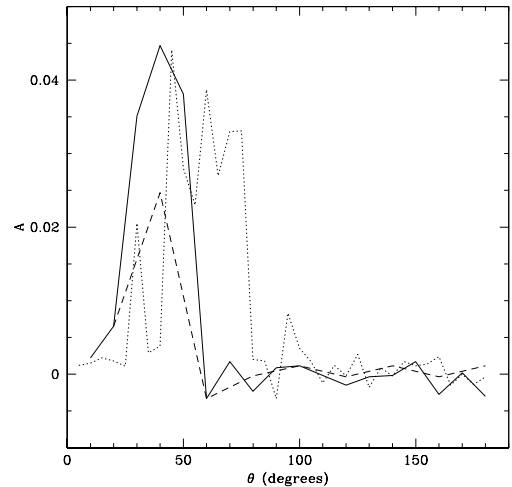}
\caption{\label{fig:epsart} Upper panel: example of anisotropy signal detection for mock fields consisting of
Hernquist spheroids plus Poisson background. Lower panel: detection of anisotropy using three angular steps:
$\Delta\theta=5^\circ,10^\circ,~{\rm and}~20^\circ$, in dotted, solid, and dashed lines, respectively.}
\end{figure}

The algorithm approximates the point pattern and its window with
binary pixel images, introduces a Gaussian smoothing kernel, and
uses the Fast Fourier Transform to form a density estimate $\kappa$. 
The calculation takes into account the edge correction known
as the ``translation correction'' (see Ripley 1977).
The density estimate of $\kappa$ is returned in the form of a
real-valued pixel image. 
The $\hat{K}$ estimator is defined as the expected number of points 
lying within a distance $r_{max}$ of a typical point, and with a displacement 
vector of orientation in the range $[\alpha, \beta]$. 
This can be computed by summing the entries over the relevant region, i.e, the
sector of the disc of radius $r$ centred on the origin with angular range
$[\alpha,\beta]$. Hence, we can compute a measure of anisotropy (A) as 
integrals of the 
form 

\begin{equation}
A\equiv\int_0^{r_{max}}\int_\alpha^\beta d\kappa(r,\theta).
\end{equation}

Note that the second-order moment function K is used to
test the hypothesis that a given planar point pattern is a realization
of a Poisson process. Thus, the objective is to
search for significant peaks in the integral given in Eq.(1) for angular steps 
$\Delta\theta=\beta-\alpha$. The uncertainty
in the anistropy measurement is computed directly from the variance
in the pixel values in the corresponding circular sector.
Of course, the choice of $\Delta\theta$ is arbitrary. In this work, we
assume that $\Delta\theta=10^\circ$, based on the
following analysis. 

Consider a controlled sample corresponding to a point pattern
given by a Poisson distribution (100 points) plus a central group defined
as a Hernquist spheroid (Hernquist 1990). An additional 
group is  located initially at $\sim 4 h^{-1}$ Mpc around an angle of $45^\circ$,
and then around an angle of $135^\circ$.
In Fig. 1, we present both the mock field and the anisotropy signal
as a function of the angle for the two positions of the second group.
It is clear from this figure that the second group produces 
significant anisotropy signal. To justify our choice of 
$\Delta\theta$, we present in Figure 1 (lower panel)
a reanalysis in the case of the second group at $45^\circ$ for
$\Delta\theta=5^\circ,10^\circ,~{\rm and}~20^\circ$.
For $\Delta\theta=5^\circ$, we still detect the peak, but 
there are now secondary peaks and a more noisy behaviour for 
$A$. For  $\Delta\theta=20^\circ$,
the peak is still there, but now less significant.
This result suggests that in the limit of a too small value $\Delta\theta$, we have
a noisy anisotropy curve (possibly with false peaks), while
in the limit of very large $\Delta\theta$, the signal can be completely lost.
In this work, we set $\Delta\theta=10^\circ$ as
a confidence scanning angle to probe anisotropy in galalxy fields
of our sample.


\subsection{The 2dF sample}

We apply the anisotropy estimator to a sample consisting of
32  galaxy groups previously identified by Tago et al. (2006) applying the
friends-of-friends algorithm to data from the 2dFGRS (Colless et al. 2001). 
This subset corresponds to those groups located in areas
of at least 80\% redshift coverage out to 10 times the
virial radius roughly estimated from the projected harmonic mean.
Group members and interlopers were redefined after the
identification of gaps in the redshift distribution according
to the technique described by Lopes et al. (2009). 
Before selecting group members and rejecting interlopers we first
refine the spectroscopic redshift of each group and identify its
velocity limits. For this purpose, we employ the gap-technique described 
in Katgert et al. (1996) and Olsen et al. (2005) to identify 
gaps in the redshift distribution. A variable gap, called 
{\it density gap} (Adami et al. 1998), is
considered. To determine the group redshift, only galaxies within 
0.50 h$^{-1}$ Mpc are considered. Details about this procedure are found 
in Lopes et al. (2009). 

\begin{figure}
\centering
\includegraphics[width=7 cm,height=7cm]{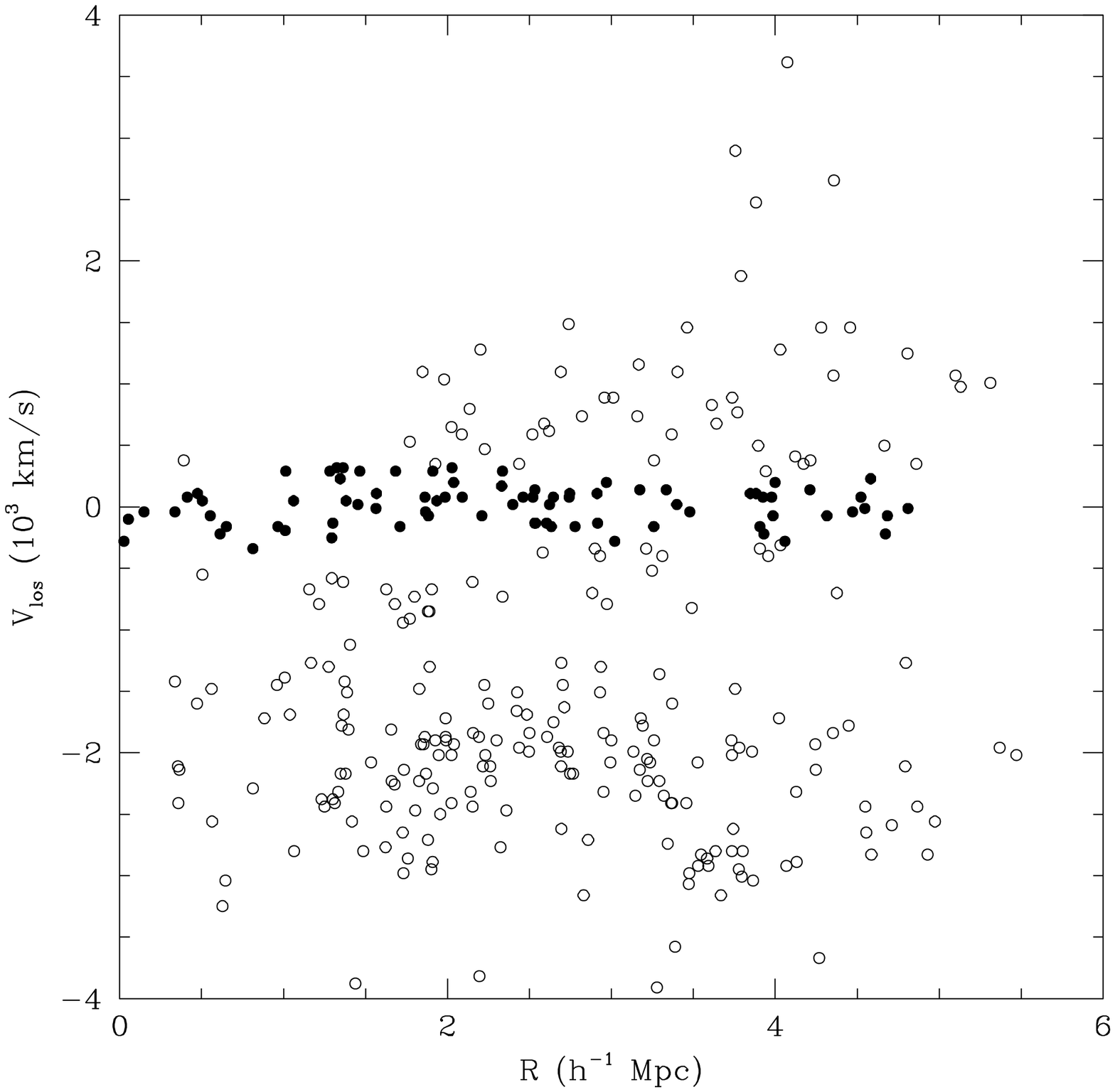}
\includegraphics[width=7 cm,height=7cm]{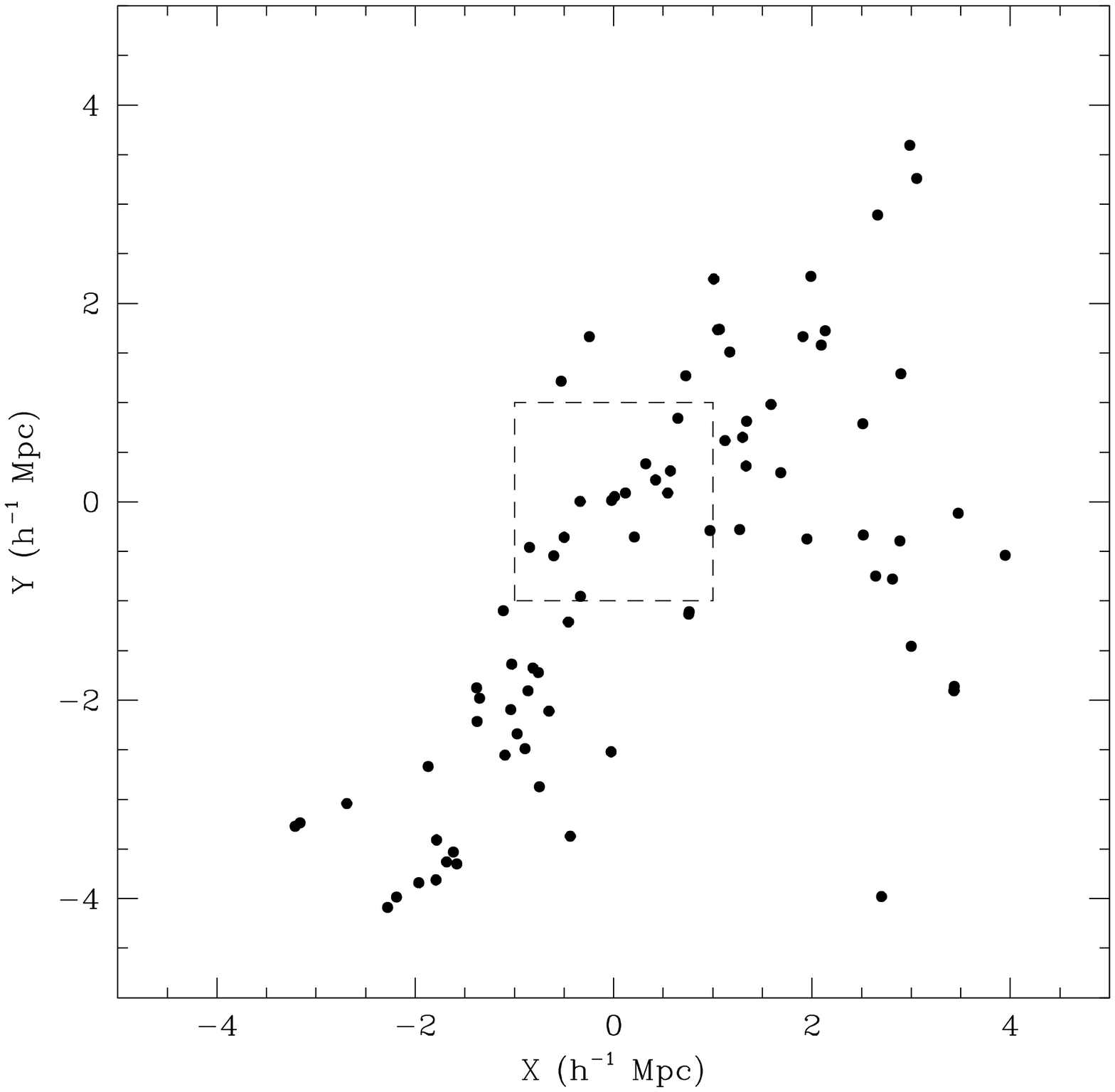}
\caption{\label{fig:epsart}Upper panel: phase-space diagram of Group 91 shown 
as an example. We consider the group center to derive the velocity and radial 
offsets. Group members (filled circles) are selected with a shifting gapper 
procedure. The interlopers are represented by open circles. 
Lower panel:  Group members with a central 1 ${\rm h^{-1}}$ Mpc square in
dashed lines.}
\end{figure}

With the new redshift and velocity limits, we apply an algorithm for interloper 
rejection to define the final list of group members. We use the
``shifting gapper'' technique (Fadda et al. 1996), which consists of the 
application of the gap-technique to radial bins from the group center.
We consider a bin size of 0.42 h$^{-1}$ Mpc (0.60 Mpc for h = 0.7) or 
larger to ensure that at least 15 galaxies are selected. Galaxies not 
associated with the main body of the group are discarded. This procedure 
is repeated until the number of group members is stable and no further galaxies 
are eliminated as intruders. An example of the application of the shifting gapper procedure
is seen in Fig. 2 (upper panel). The main difference from the study of Lopes et al. (2009) is that here
we consider all galaxies within 10 times the virial radius (as listed in
Tago et al. 2006). In Lopes et al. (2009), the interloper removal procedure 
was applied to galaxies within a maximum radius of 2.5 h$^{-1}$ Mpc. Next, 
we estimate the velocity dispersions ($\sigma$) 
and physical radius (R$_{200}$) of each group. Finally, a virial 
analysis is perfomed for mass estimation (M$_{200}$). Further 
details regarding the interloper removal and estimation of global properties 
($\sigma$, physical radius and mass) are found in Lopes et al. (2009). 

\begin{figure}
\includegraphics[width=8.5cm,height=8.5cm]{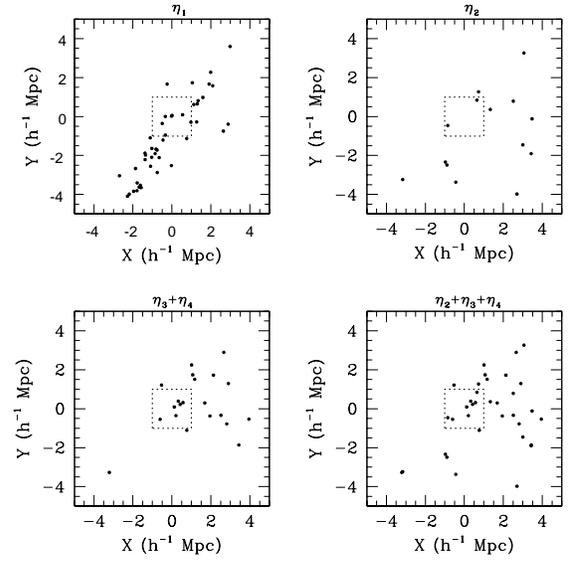}
\caption{\label{fig:epsart} Galaxy distribution around Group 91 per $\eta$-type division with central 1 ${\rm h^{-1}}$ Mpc squares in
dashed lines.}
\end{figure}

The physical properties of these groups are presented
in Table 1. The columns correspond to:


\begin{enumerate}
\item Group identification number;
\item RA (J2000.0) in degrees (mean of member galaxies);
\item DEC (J2000.0) in degrees (mean of member galaxies);
\item  z, the new redshift, determined within 0.5 $h^{-1}$ Mpc;
\item  $\sigma$, the velocity dispersion in km${\rm s^{-1}}$ (computed with the group members);
\item  ${\rm R_{200}}$ in Mpc;
\item  ${\rm M_{200}}$ (${\rm 10^{14} ~M_\odot}$);
\item  Number of member galaxies (after exclusion of interlopers);
\item  Number of member galaxies within $R_{200}$;
\item  Global anisotropy (see definition in Section 2.3);
\item  Fraction of $\eta_1$ galaxies (Section 2.3);
\item  Fraction of galaxies within $R_{200}$.
\end{enumerate}

\begin{table*}
\caption{Properties of groups}             
\label{table:1}      
\centering                          
\begin{tabular}{r c c c c c c c c c c c}        
\hline\hline                 
Group & RA & DEC &  z & $\sigma$  & $R_{200}$ & $M_{200}$ & 
$N$ & $N_{200}$& GA & $f_{\eta_1}$& $f_{200}$  \\    
    &  ($\circ$) & ($\circ$) & & (km ${\rm s^{-1}}$) & (Mpc) & (${\rm 10^{14} ~M_\odot}$) & &  & & \\
\hline                        
  23 & 168.608 & -4.008 & 0.1010 & 264.202 & 0.65 & 0.35 & 16 & 13 & 2.894 & 0.645 & 0.812\\
  59 & 162.581 & -0.332 & 0.0949 & 221.623 & 0.55 & 0.21 & 13 & 10 & 3.263 & 0.687 & 0.769\\
  60 & 162.231 &  0.936 & 0.1070 & 262.031 & 1.28 & 2.66 & 65 & 18 & 2.939 & 0.500 & 0.276\\
  61 & 161.688 &  1.438 & 0.1068 & 290.062 & 1.20 & 2.19 & 73 & 22 & 4.152 & 0.550 & 0.301\\
  63 & 160.890 &  1.555 & 0.1065 & 198.729 & 1.03 & 1.37 & 46 & 10 & 2.992 & 0.520 & 0.217\\
  64 & 160.419 &  1.361 & 0.0726 & 169.913 & 0.79 & 0.61 & 43 & 14 & 2.903 & 0.625 & 0.325\\
  83 & 174.909 & -1.110 & 0.0777 & 236.053 & 0.99 & 1.20 & 54 & 20 & 2.746 & 0.541 & 0.370\\
  86 & 180.818 &  0.882 & 0.0782 & 125.600 & 0.66 & 0.35 & 36 & 12 & 2.357 & 0.566 & 0.333\\
  91 & 191.534 &  0.703 & 0.0892 & 155.980 & 0.89 & 0.88 & 80 & 9  & 3.483 & 0.562 & 0.112\\
  95 & 191.946 & -0.213 & 0.0898 & 219.657 & 0.95 & 1.06 & 58 & 20 & 2.409 & 0.541 & 0.344\\
 137 &  23.602 & -32.823& 0.0646 & 362.672 & 1.08 & 1.51 & 45 & 25 & 3.690 & 0.500 & 0.555\\
 139 &  32.829 & -33.200& 0.1065 & 179.301 & 0.84 & 0.75 & 34 & 12 & 3.235 & 0.604 & 0.352\\
 182 &   6.925 & -30.745& 0.1066 & 279.339 & 1.34 & 3.01 & 74 & 18 & 2.553 & 0.479 & 0.243\\
 187 &  10.290 & -29.003& 0.1089 & 165.089 & 0.90 & 0.93 & 59 & 12 & 2.654 & 0.562 & 0.203\\
 188 &  14.169 & -30.784& 0.0769 & 256.686 & 1.16 & 1.94 & 64 & 17 & 1.917 & 0.500 & 0.265\\
 189 &   9.426 & -30.787& 0.0613 & 249.968 & 0.93 & 0.98 & 36 & 15 & 3.259 & 0.541 & 0.416\\
 193 &  17.559 & -29.663& 0.1074 & 151.658 & 0.82 & 0.71 & 46 & 11 & 3.724 & 0.604 & 0.239\\
 195 &  15.688 & -31.925& 0.1087 & 222.339 & 0.98 & 1.21 & 40 & 16 & 2.759 & 0.541 & 0.400\\
 208 &  36.363 & -30.062& 0.0723 & 270.511 & 0.82 & 0.68 & 26 & 16 & 2.194 & 0.625 & 0.615\\
 216 &  47.233 & -31.051& 0.0650 & 222.393 & 1.09 & 1.55 & 59 & 16 & 3.593 & 0.550 & 0.271\\
 225 & 328.094 & -28.863& 0.0929 & 197.640 & 0.97 & 1.14 & 45 & 17 & 2.517 & 0.583 & 0.377\\
 245 & 352.361 & -30.148& 0.1057 & 378.582 & 1.60 & 5.16 & 58 & 12 & 2.168 & 0.479 & 0.206\\
 249 & 355.576 & -30.284& 0.0619 & 234.400 & 1.01 & 1.24 & 45 & 24 & 3.115 & 0.562 & 0.533\\
 250 & 355.214 & -30.339& 0.0801 & 246.493 & 0.71 & 0.43 & 22 & 14 & 1.855 & 0.645 & 0.636\\
 256 &  10.190 & -26.482& 0.1121 & 342.401 & 1.31 & 2.83 & 57 & 15 & 3.091 & 0.479 & 0.263\\
 258 &   9.870 & -26.520& 0.1009 & 158.659 & 0.66 & 0.36 & 19 &  8 & 3.615 & 0.666 & 0.421\\
 264 &  26.607 & -26.639& 0.0600 & 301.131 & 0.98 & 1.14 & 22 & 13 & 3.717 & 0.604 & 0.590\\
 266 &  32.245 & -26.366& 0.1162 & 271.840 & 0.83 & 0.72 & 17 & 10 & 2.157 & 0.562 & 0.588\\
 269 &  42.215 & -26.001& 0.1052 & 239.011 & 1.05 & 1.47 & 39 &11  & 2.986 & 0.500 & 0.282\\
 278 & 334.293 & -26.673& 0.0598 & 315.139 & 1.07 & 1.47 & 37 &15  & 2.351 & 0.520 & 0.405\\
 280 & 343.716 & -25.911& 0.0805 & 168.699 & 0.74 & 0.51 & 23 &12  & 2.690 & 0.566 & 0.521\\
 281 & 343.461 & -25.517& 0.0899 & 234.579 & 1.06 & 1.48 & 48 &19  & 2.711 & 0.541 & 0.395\\
\hline                                  
\end{tabular}
\end{table*}




\subsection{Description of one group + surroundings field}

The methodology presented in Sect. 3.1 is now applied to describe in detail
the anisotropy features of Group 91 plus its neighborhood. 
This field is presented in Fig. 2 (lower panel), where equatorial coordinates were transformed to
Cartesian ones using redshift information for a
flat universe with $\Omega_m=0.3$. We can see that the galaxy distribution is 
clearly anisotropic. In Fig. 3, we present the same field, but
now we identify differences in the general behaviour according to the galaxy spectral type defined by
Madgwick et al. (2002). The $\eta$ parameter is a measure of
spectral type, which  corresponds approximately to the following division:

\hspace{0.2 cm}

\framebox[5.25cm][c]{%
$\eta_1 \longrightarrow$ E-S0 galaxies $[\eta < -1.4]$~~~~~~} \par
\framebox[5.25cm][c]{%
$\eta_2 \longrightarrow$ Sa galaxies $[-1.4\le \eta < 1.1]$~}\par
\framebox[5.25cm][c]{%
$\eta_3 \longrightarrow$ Sb galaxies $[1.1\le \eta < 3.5]$~~~}\par
\framebox[5.25cm][c]{%
$\eta_4 \longrightarrow$ Scd galaxies $[\eta\ge 3.5]$~~~~~~~~~~~}

\hspace{0.2cm}

\noindent (see Madgwick et al. (2002)  for more details about the spectral type classification and
data division by $\eta$).

In this work, we ascertain whether the anisotropy pattern is related to the
galaxy types in the group + surroundings field. The anisotropy profile of Group 91 is presented in Fig. 4 (upper panel),
where we see the overall behavior in the small box and the behavior per type
in colors (main box). Typically, our fields are so dominated by $\eta_1$ galaxies 
($\sim 60\%$ in average)  that
the number of objects in the remaining bins are too small for comparisons
between each other. Here, we 
just compare $\eta_1$ with the other types, according to 
$\eta_1 \times \eta_2$, $\eta_1 \times (\eta_3+\eta_4)$, and
$\eta_1 \times (\eta_2+\eta_3+\eta_4)$. To verify whether the respective profiles
differ significantly from each other, we applied a bootstrap hypothesis test, assuming as the null hypothesis that the mean 
of the difference in the anisotropy signal between two 
populations is zero ($H_0:\mu=0$), while the alternative 
hypothesis is $\mu\neq 0$.
A small $p$-value makes the null hypothesis appear implausible. Obtaining $p$ can be done using
the bootstrap resampling approach. In this work, based on 1,000 bootstrap sample replications, we 
obtain for Group 91:  $\eta_1 \times \eta_2 :$ p=0.008;
$\eta_1 \times \eta_{34} :$ p=0.014; and
$\eta_1 \times \eta_{234} :$ p=0.015.
Taking the usual cutoff of $p=0.05$ (5\% significance level), we reject the null hypothesis
in all cases and conclude that $\eta_1$-type galaxies in this case have a distinctive spatial distribution.
Additionally, determining the group + surroundings shape by diagonalizing the moments of 
the inertia tensor as a function of the radius, we find that there is an elongation jump close to $R_{200}$ (see Fig. 4, lower panel). 
This suggests that the group is embedded in a highly anisotropic structure.

We applied this methodology to the remaining groups
of our sample and found that only 15\% of the groups (91,139,189,193, and 250) have a distinctive $\eta_1$ population.
However, the number of fields with significant elongation is quite high. To help us understand that, we define a new quantity, the global anisotropy (GA) as

\begin{equation}
{\rm GA={Max(A) - Median(A)\over \sigma(A)}}.
\end{equation}

\noindent Basically, this is a measure of the relative importance of the 
anisotropy peak with respect to the entire profile. The values of GA are listed in Table 1. Here, we require that a significant elongation
corresponds to GA $\ge$ 2, i.e., Max(A) $\ge$ 2$\sigma$(A) + Median(A). Following this criterium, we
have found that 94\% of the fields have a high degree of elongation. Being more restrictive and setting GA$\ge$3,
37\% of the systems still have this aligment prominence. In our control sample of 1,000 Hernquist spheroids + Poisson background, we
found just 12\% of aligments by chance. Hence, we conclude that our sample consists of a significant number of group + surroundings 
predominantly elongated.



\begin{figure}
\centering
\includegraphics[width=7 cm,height=7cm]{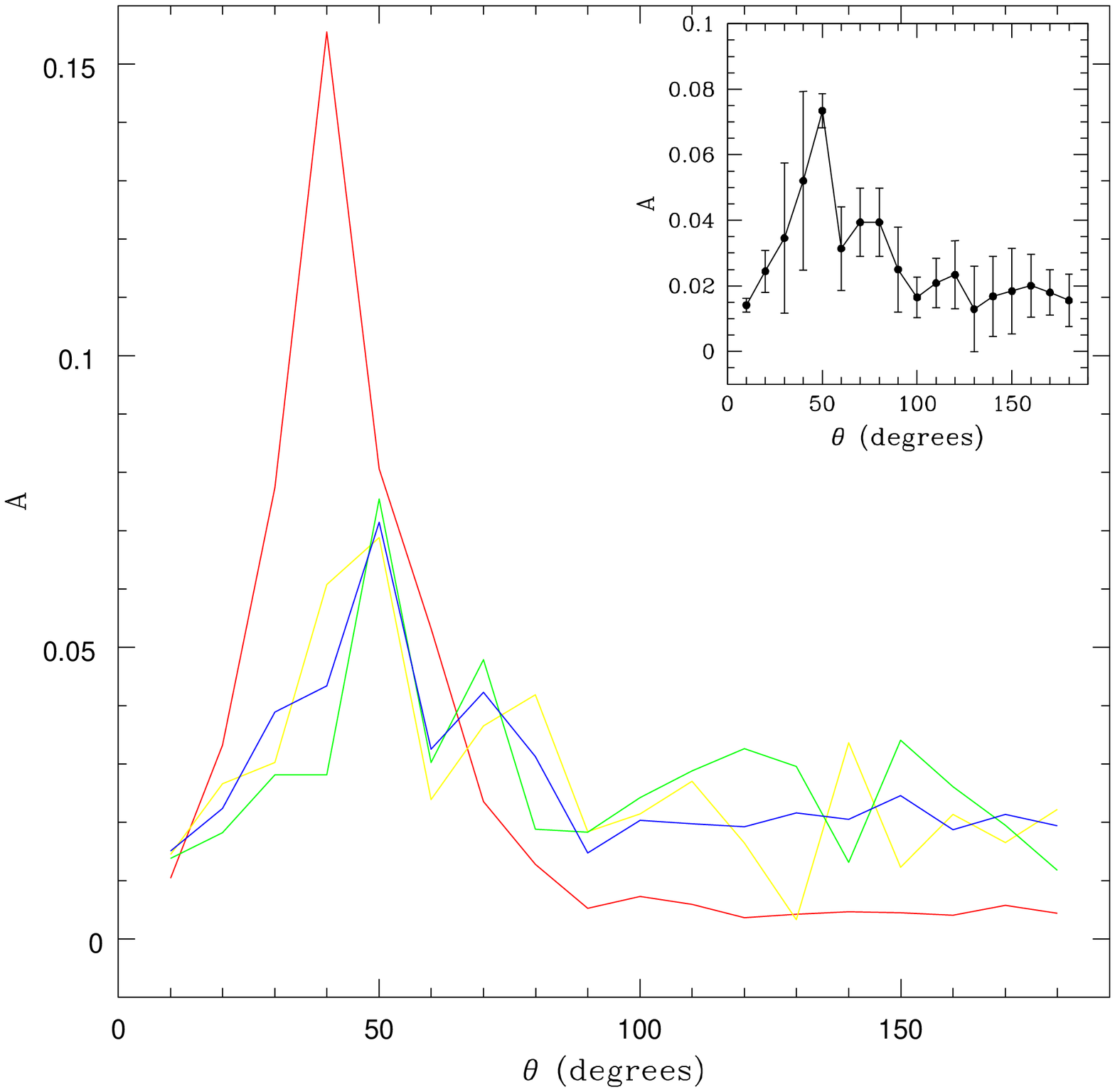}
\includegraphics[width=7 cm,height=7cm]{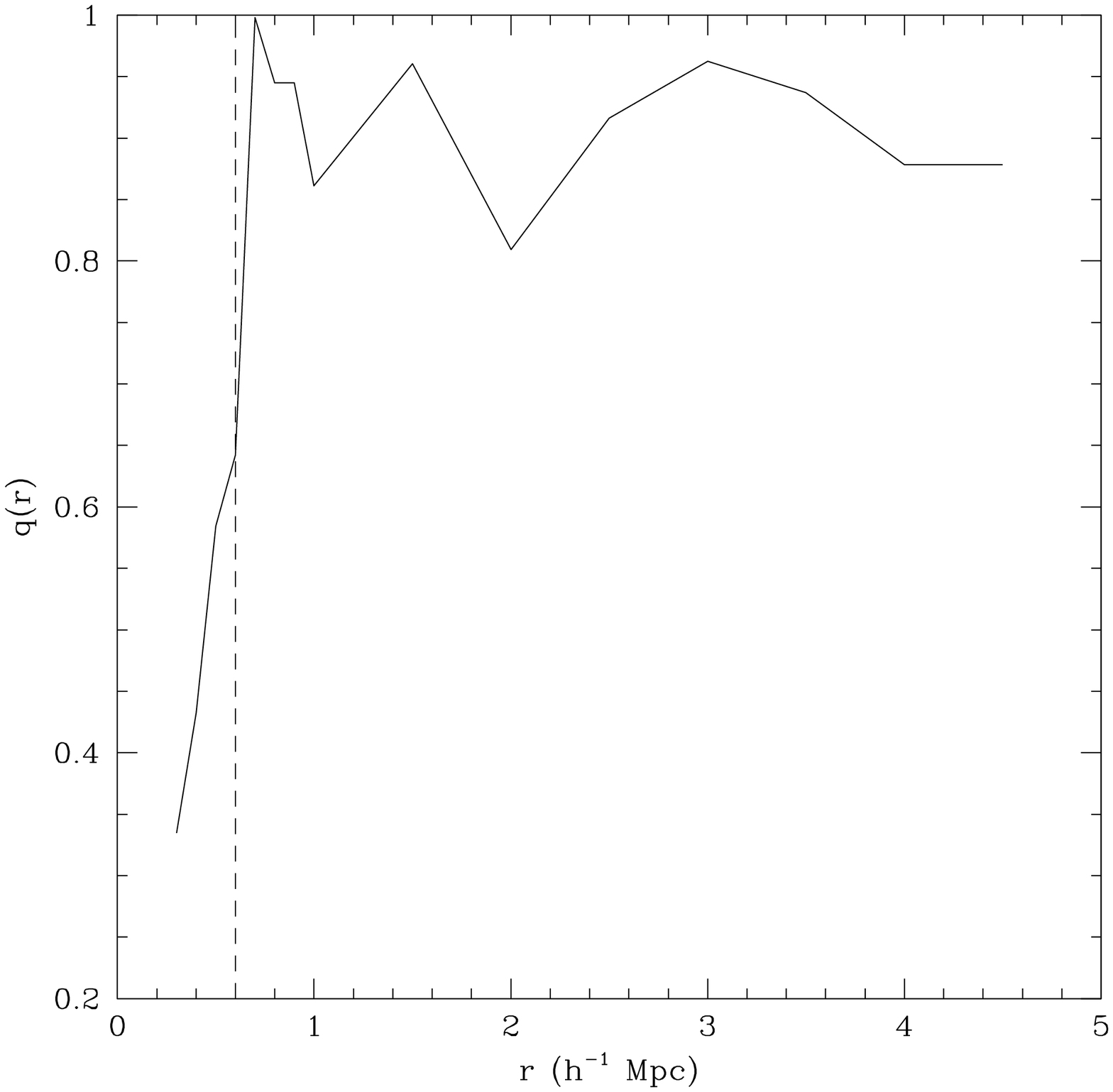}
\caption{\label{fig:epsart} Upper panel: anisotropy profile of Group 91. $\eta_1$ galaxies are in red; $\eta_2$ galaxies are in yellow;
$\eta_3+\eta_4$ galaxies are in green; and $\eta_2+\eta_3+\eta_4$ galaxies are in blue. The general behaviour is displayed in
the small box. Lower panel: ratio of eigenvalues $q=\lambda_1/\lambda_2$ as a function of the radius, obtained after
diagonalizing the inertia tensor $\det{(I_{ij} - \lambda^2M)=0}$ (where M is a $2\times 2$ unit matrix). The vertical
dashed line indicates $R_{200}$. }
\end{figure}


\section{Anisotropy and galaxy evolution}

The intrinsic elongated shape of groups can be a very important factor when determining their
dynamical state (e.g. Tovmassian \& Plionis, 2009). In this section, we investigate a possible correlation
between anisotropy and galaxy evolution for all galaxies in our sample. 

\subsection{Characterizing galaxies around groups}

We use the
spectral parameter $\eta$ to characterize galaxy evolution around groups.
We applied the Welch's $t$-test (Welch 1947),
also known as the $F$-test (Press, Flannery \& Teukolsky 1986), which is an adaptation
of the Student's $t$-test, to the comparison of two samples (in this case $\eta< -1.4$ and $\eta \ge -1.4$) with unequal variance. This statistic tests the null hypothesis that
the mean of each of the two samples are equal, assuming a normally distributed parent population.
The test was applied in two ways: (i) testing the two samples by keeping the
divisory line $\eta=-1.4$; (ii) testing data iteratively in samples defined by one of the
other quantities ($M_B$, $B-R$, and distance to the center of the groups, $d_c$). 
In the second case, the two data sets are redefined for increasing (or decreasing) values of the quantity, so we can find the corresponding divisory line (i.e., the first value in which
the null hypothesis is rejected).

We first probe the distribution
of galaxies as a function of the distances to the center of the groups.
To illustrate our results more clearly, galaxies were sorted in $\eta$ and
divided into seven subgroups with the same number of objects.
In Fig. 5 (upper panel), we see that low $\eta$
(early type) galaxies are more concentrated than high $\eta$ objects. 
We verified a corresponding horizontal line at $d_c/R_{200}=1.5$,
where data can be divided into two statistically distinct groups, after a $t$-test ($p=1.78\times 10^{-5}$). 
This expected result is just a manifestation of the morphology-density (radius) relation (e.g., Dressler 1980).
We also find that our sample is dominated by dwarf galaxies ($M_B\ge -20$
for $\sim$90\% of the sample) and that low $\eta$ galaxies
are more luminous than the remainder (see Fig. 5, middle panel), where a horizontal line at $M_B=-18.55$ divides data
into two subgroups that are statistically distinct ($p=1.02\times 10^{-7}$). Finally, the distribution of (B-R) color indicates
two distinct groups at B-R=1.1 ($p=8.07\times 10^{-5}$), where low $\eta$ galaxies are redder than the rest 
(see Fig. 5, lower panel). Thus, our sample is dominated 
by dwarfs, and low $\eta$ objects are more central, luminous and redder than the other galaxies.

\begin{figure}
\centering
\includegraphics[width=6 cm,height=6cm]{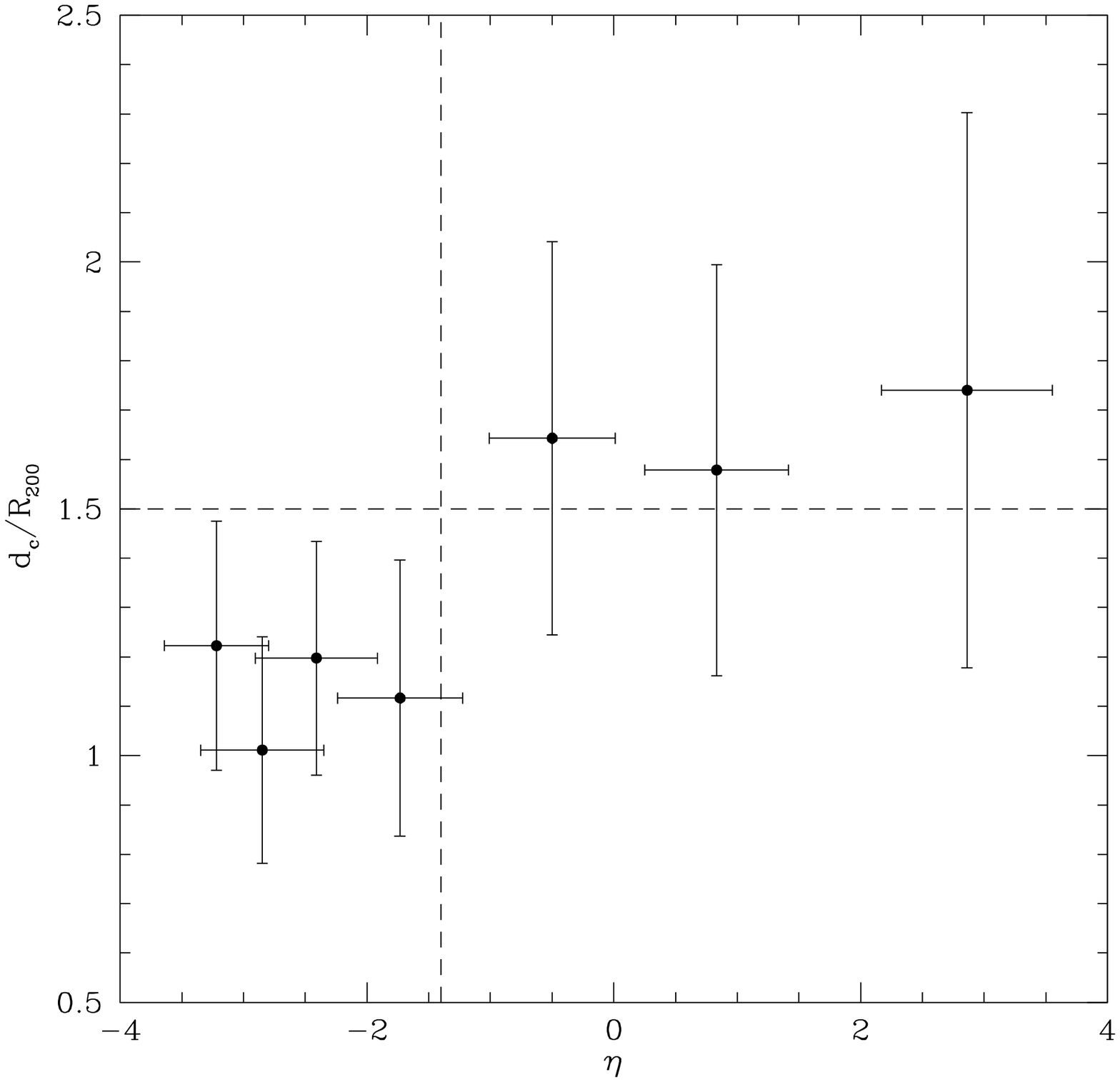}
\includegraphics[width=6 cm,height=6cm]{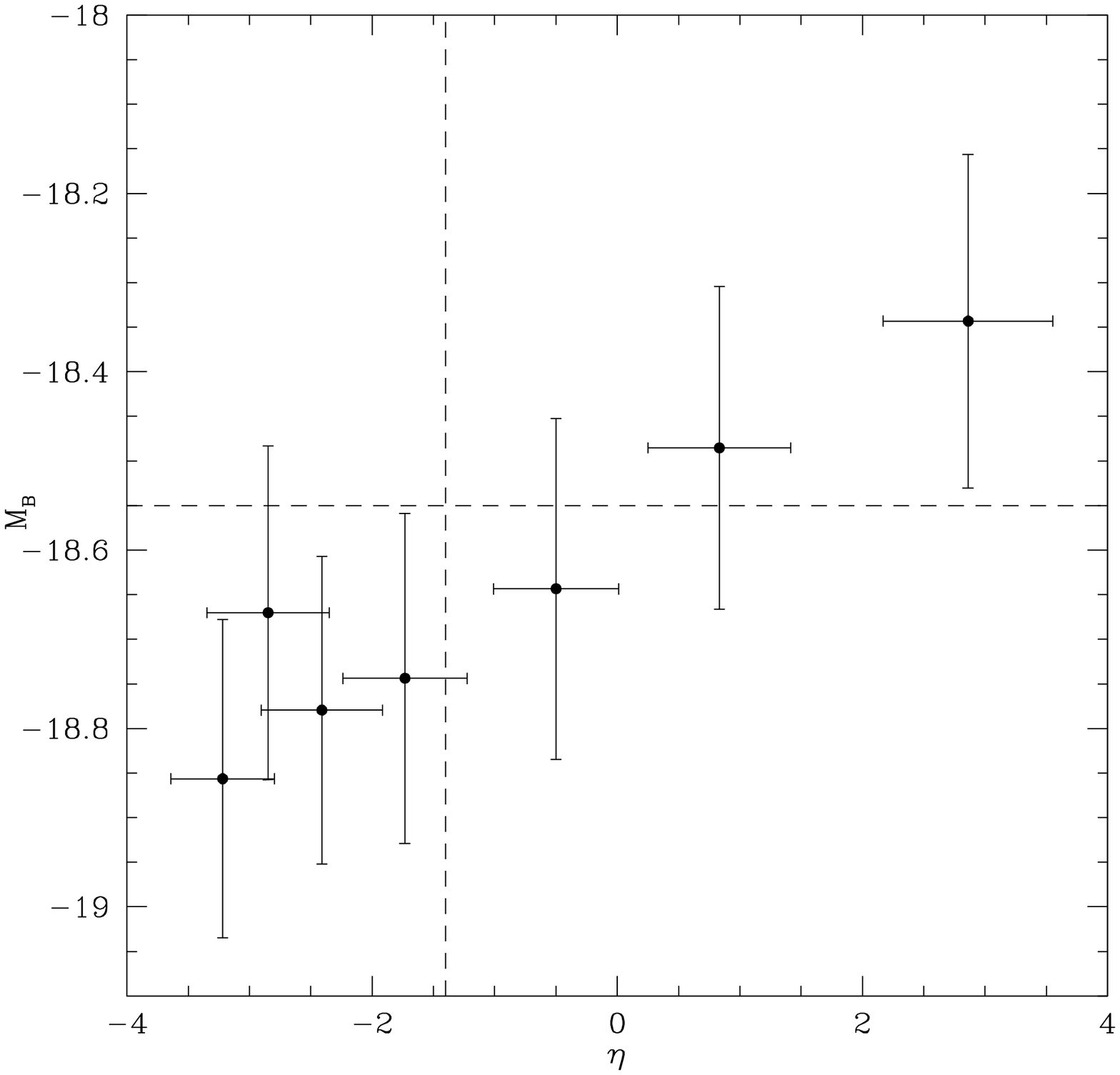}
\includegraphics[width=6 cm,height=6cm]{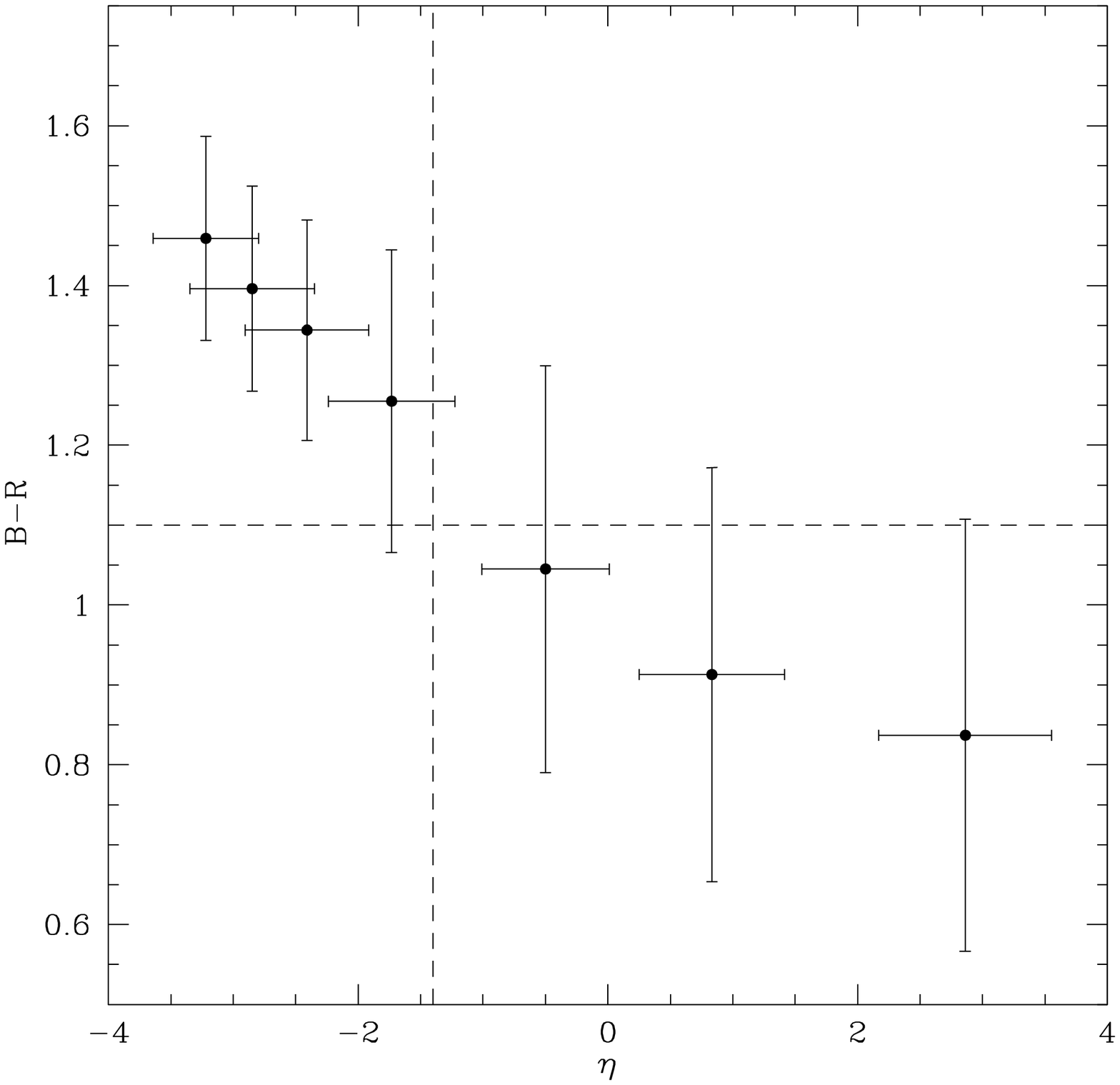}
\caption{\label{fig:epsart} Distances of galaxies to the center of the groups (normalized by
$R_{200}$) as a function of $\eta$ (upper panel). Absolute magnitudes as a function of $\eta$ (middle panel). B-R color as a function of $\eta$ (lower panel). 
Dashed lines divide data into two groups statistically distinct.}
\end{figure}

\subsection{Global anisotropy, galaxy evolution, and dynamics}

Now, we wish to study the possible correlation between galaxy evolution and anisotropy.
We verified  that the fraction of $\eta_1$ galaxies around groups exhibits a significant correlation with GA. This result, presented
in Fig. 6 (lower panel), suggests that a relation existis between the shape of groups + surroundings, and their
star formation rate. More anisotropic fields have a lower fraction of $\eta_1$ galaxies
(the $F$-test presents $p=0.0036$ for a linear fit).  
We also divide galaxies into GA quartiles and probe
their distribution with distance to the center. This is also presented in Fig. 6 
(upper panel), where we see no significant difference between the GA quartiles
with increasing radius. The general behaviour just reproduces the morphology-density (radius) relation.

\begin{figure}
\centering
\includegraphics[width=7.0 cm,height=4.5cm]{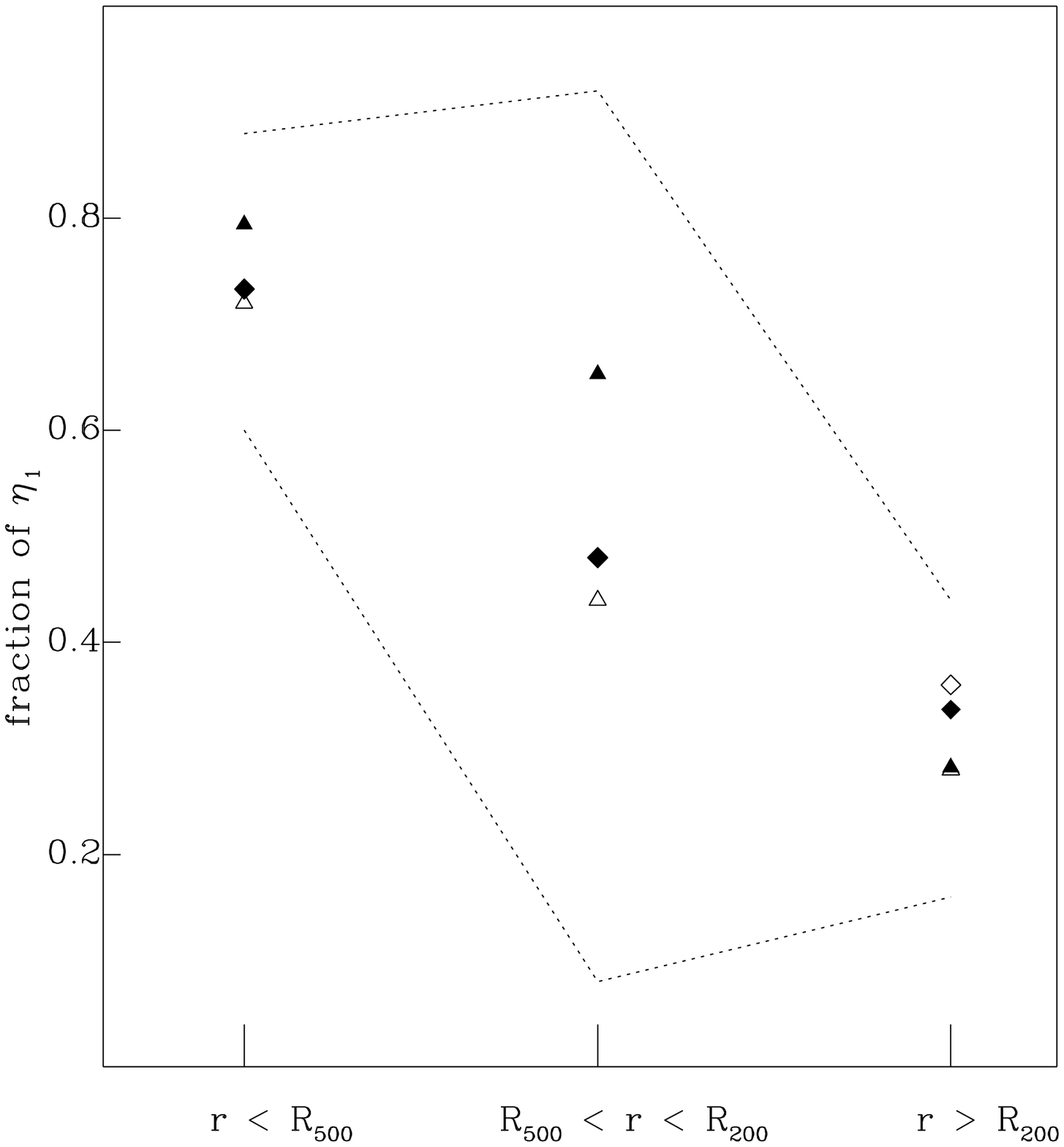}
\includegraphics[width=6.8 cm,height=6.5cm]{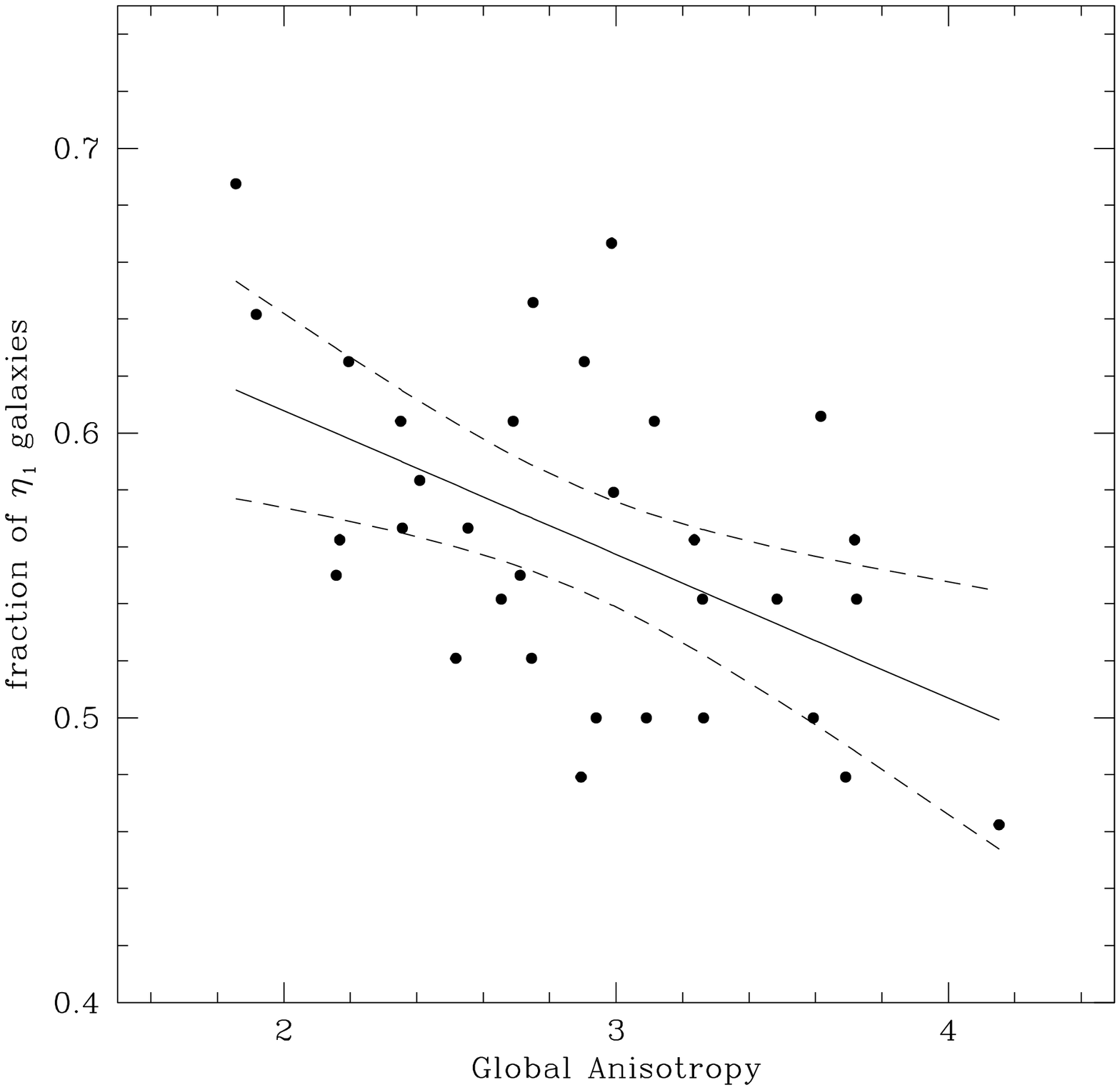}
\caption{\label{fig:epsart} Lower panel: global anisotropy versus the fraction of $\eta_1$ galaxies. Best fit linear is in solid lines, while 95\% confidence levels contours are presented as dashed lines. Upper panel: distance against the fraction of $\eta_1$ galaxies, now divided in GA quartiles (open and filled losangles, open and filled triangles). Dotted lines
show the typical errors for each distance range.}
\end{figure}

The $\eta_1$ galaxy fraction in the fields is a good indicator of evolution, since
the morphology-density relation appears to imply that late become early type galaxies.
We now present some additional trends in our data associated to 
this quantity. In Fig. 7, we see that $f_{\eta_1}$ is anticorrelated
 with $\sigma$ (p= 0.0066) and $N$ (p=2.8$\times 10^{-5}$). That is, cold groups evolving in poorer environments
contain a higher fraction of $\eta_1$ galaxies. At the same time, $f_{\eta_1}$ is correlated with $f_{200}$ (p=0.0002):
groups in which more galaxies are inside $R_{200}$ exhibits a higher fraction of $\eta_1$ galaxies (Fig. 7). 
Finally, $f_{200}$ is anticorrelated with $N$ (p=1.8$\times 10^{-8}$), (see also in Fig. 7). This all means that
rich fields harbour less concentrated galaxy systems with fewer ${\eta_1}$ galaxies, i.e., less evolved groups. These fields 
are also more anisotropic, containing hotter groups, which is more consistent with 
a scenario where galaxies move along the elongation direction, 
as expected in dynamical young systems that form by anisotropic accretion of matter along 
filamentary large-scale structures (e.g., Tovmassian \& Plionis, 2009). 

 \begin{figure}
\centering
\includegraphics[width=6 cm,height=6cm]{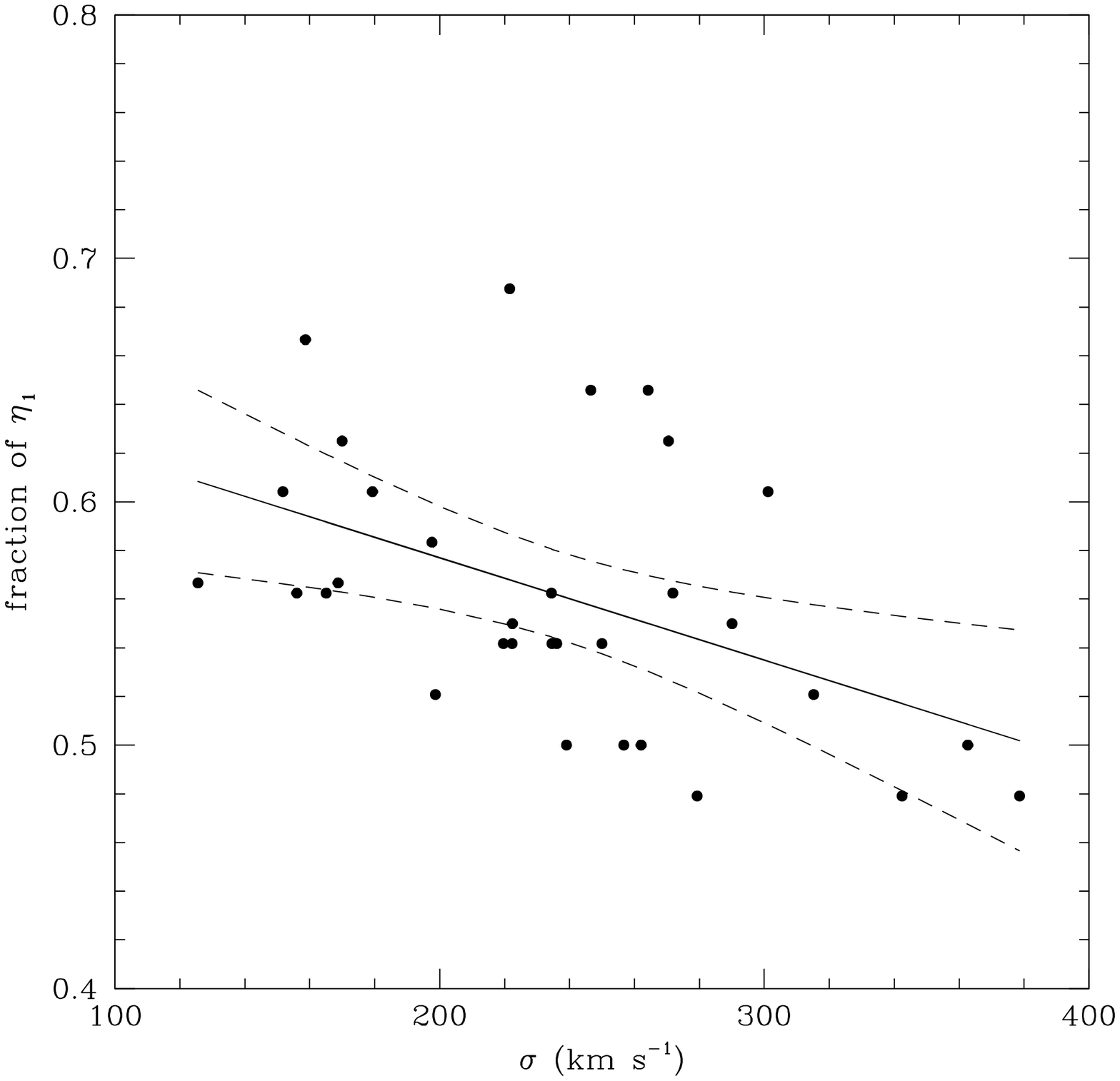}
\includegraphics[width=6cm,height=6cm]{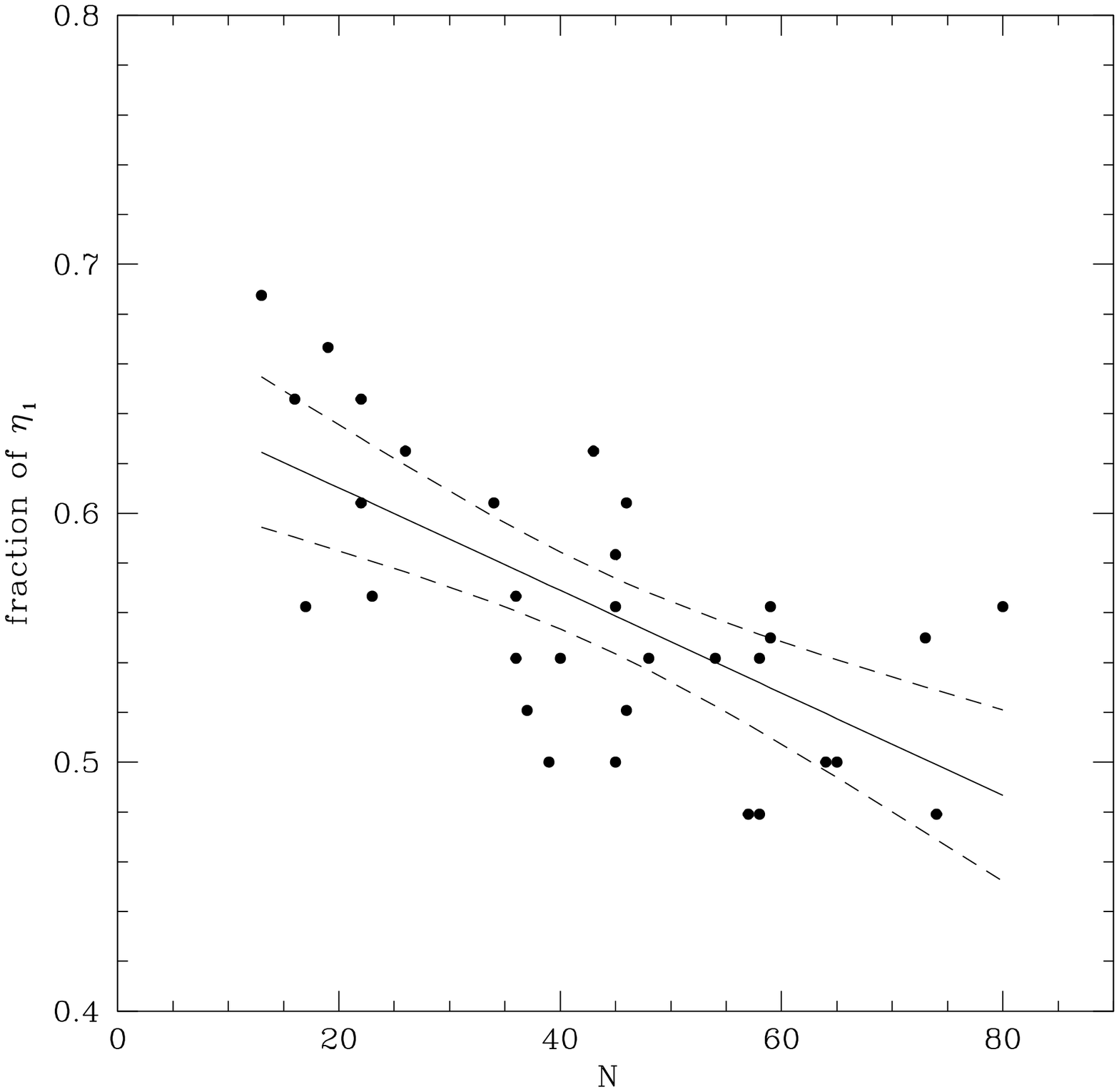}
\includegraphics[width=6cm,height=6cm]{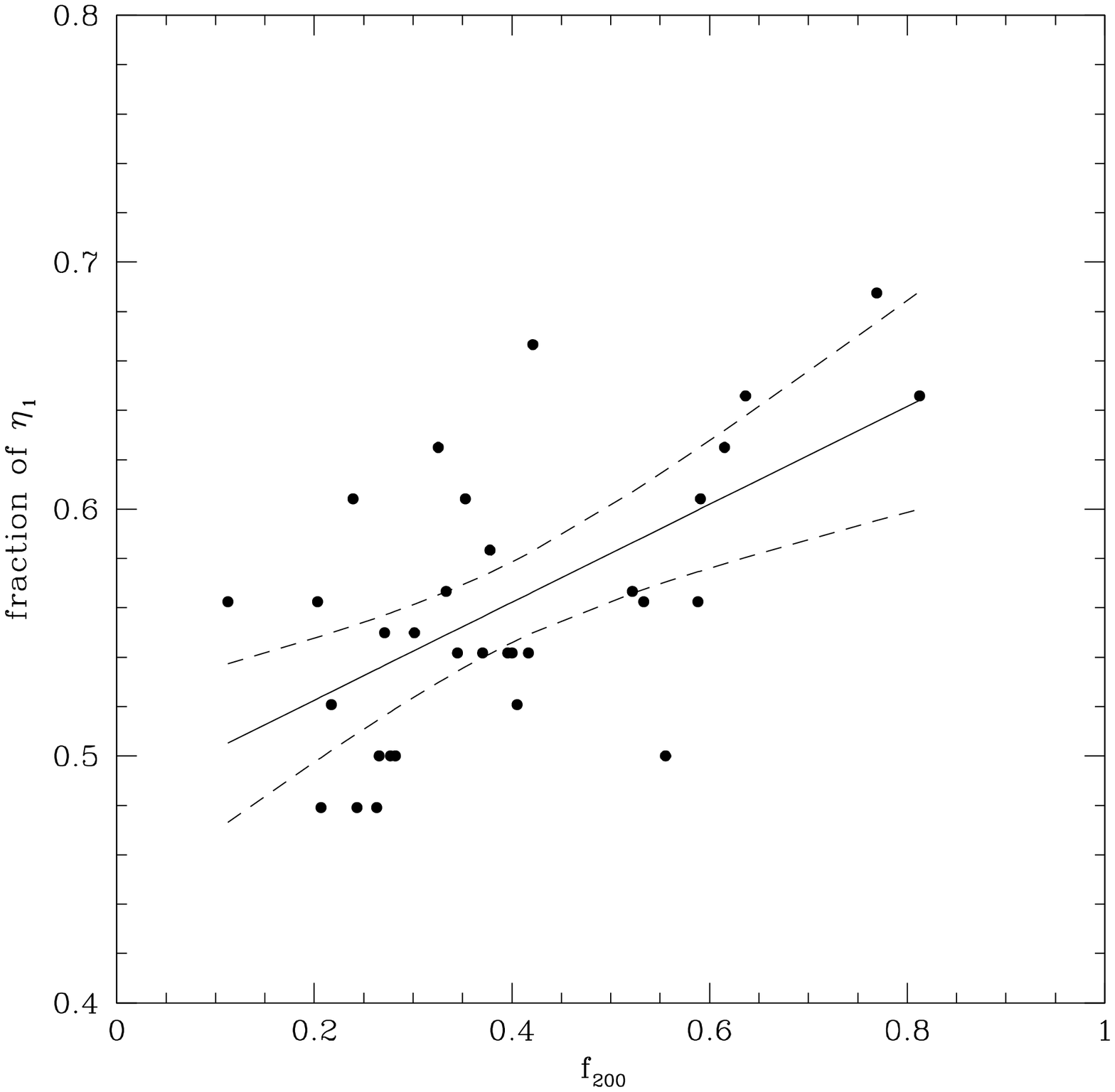}
\includegraphics[width=6cm,height=6cm]{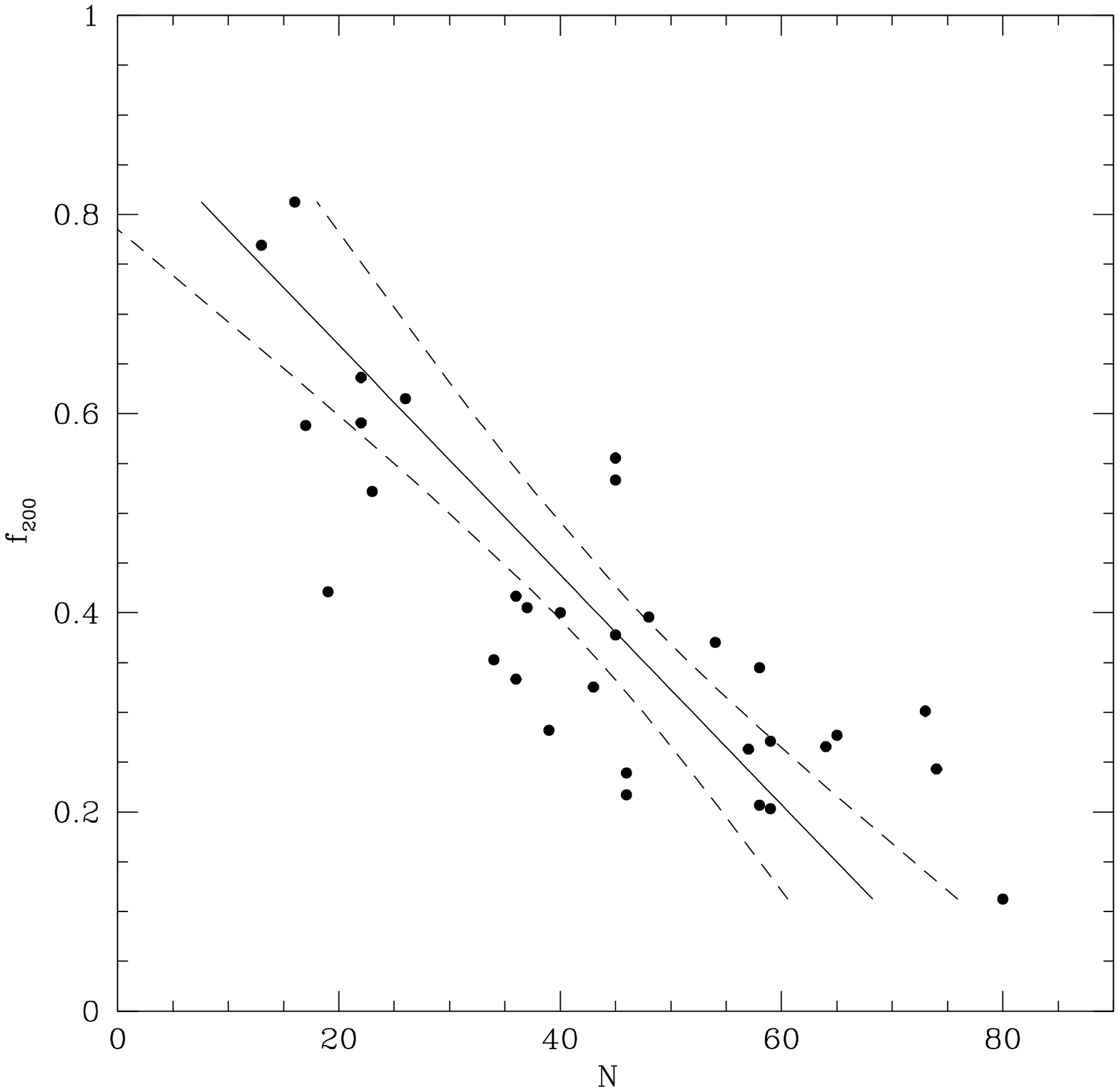}
\caption{\label{fig:epsart} Additional trends in our sample with respect the fraction of $\eta_1$ galaxies and
to the fraction of galaxies within $R_{200}$.
95\% confidence levels contours are presented as dashed lines.}
\end{figure}

Trying to extend this scenario into a more dynamical work, 
we consider the behaviour of galaxy groups in the $(\sigma,R_{200},M_{200})$ space
(see Fig. 8). We note that most of the groups settle 
onto a fitted plane given by
$M_{200} = 0.001\sigma + 3.985 R_{200} -  2.792$ (p=1.2$\times 10^{-15}$, for a F-statistic). Dividing groups according to their galactic content, i.e., the fraction of $\eta_1$ galaxies, we can see a clear
difference in the group distribution on this plane. Open circles denote groups with $f_{\eta_1}> 0.55$, which are
distributed more to the left and bottom of the plane, while
filled circles represent groups with $f_{\eta_1}\le 0.55$, predominantly located to the
right and top of the plane (0.55 is the median value for $f_{\eta_1}$).
The existence of a plane in the $(\sigma,R_{200},M_{200})$ space indicates that groups are sufficiently evolved for 
their properties to be well correlated in this dynamical frame. 
We note, however, that $f_{\eta_1}$ is related to entire fields, and
not only to groups. Hence, different loci on this plane for low and high $f_{\eta_1}$ fields indicate
a relation between the groups and their surroundings. Recall that $f_{\eta_1}$  is also anticorrelated with GA,
another field (not group) measurement. Hence, a tantalizing view of this result is that we have groups in different
dynamical states approximately along the diagonal from the top-right to the bottom-left corners of the fitted plane.
This is not exactly an evolutionary track, but an indicator of how anisotropy, galactic content, and dynamics are intimately
connected during the formation of galaxy systems.

 \begin{figure}
\centering 
\includegraphics[width=8 cm,height=8.5cm]{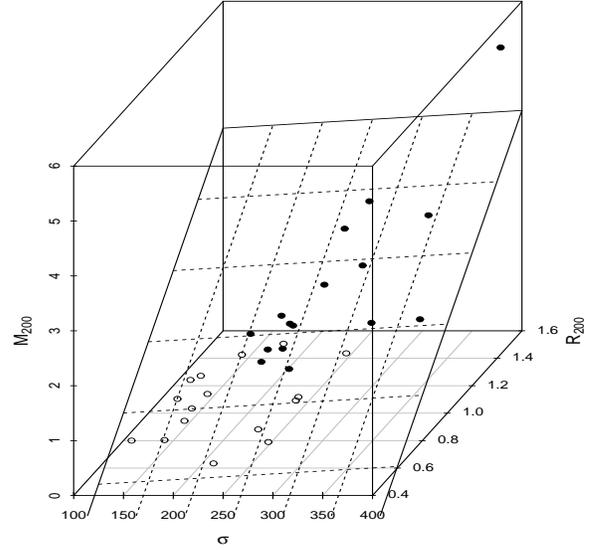}
\caption{\label{fig:epsart} 3D plot in 
$[\sigma~ ({\rm km~ s^{-1}}),R_{200}~ ({\rm Mpc}),M_{200}~ ({\rm 10^{14} ~M_\odot})]$ space. Open circles denote groups with $f_{\eta_1}> 0.55$, while
filled circles represent groups with $f_{\eta_1}\le 0.55$.}
\end{figure}

\section{Discussion}

The spatial distribution of galaxies traces the shape of the dark matter potential
in which they are embedded. Simulations show that dark matter halos are not spherical,
as one would expect from dark matter dissipationless nature, but they are strongly flattened
triaxial ellipsoids (e.g., Dubinski \& Carlberg 1991). Groups are probably the most suitable objects
to use in studying the shapes of dark matter halos, since they connect
the general field of galaxies and large-scale structure. 
Indeed, any correlation between morphological
properties of the groups and galaxy evolution can provide an indication of how matter assembles to
form larger and larger galaxy systems, and how the environment affects 
galaxy evolution during this process. 

In this work, we have introduced a new method to probe extended regions
around galaxy groups. Based on the second moment of the
Ripley function, the method allows us to define an operational anisotropy profile that
indicates preferential directions around the systems. We also define a
global anisotropy (GA) from the anisotropy profile, a quantity that can
be compared to other properties of the groups.
Galaxies in our 2dF group sample are
distributed out to $\sim 10R_V$  around the center of the groups, so we have
fairly extended samples of group+surrounding galaxies. We find that
GA is correlated  with the spectral parameter $\eta$
(Madgwick et al. 2002), an indicator of the star formation rate of galaxies.
Observations indicate that the lower star formation rate of group galaxies 
is visible out to $2R_{200}$ (Balogh et al. 1998), while
${\rm \Lambda}$CDM numerical simulations show that particles that penetrate deep into
dark matter halos travel out to $\simeq 2.6 R_{200}$ (Gill, Knebe \& Gibson 2005).
In this work, we have found that galaxies represent two statistically distinct groups
 with a transition at $\eta =-1.4$ and $d_c=1.5 R_{200}$, a scale somewhat smaller
(by 25\%) than the observed radius for decreased star formation, but consistent
with this value.
At the same time, our sample  by dwarf galaxies ($M_B>-20$ for $\sim 90\%$ of all
objects) with statistical transition line at $M_B=-18.55$, 
such that the central galaxies are the most luminous as well. 
These are also redder than the
more external ones, with a transition line at $B-R=1.1$. All of this suggests that our
sample consists mainly of dwarf ellipticals (dE). Interestingly, dE are potentially the only galaxy
type whose formation is sensitive to global, rather than local, 
environment (Conselice 2005). In this context it is important to note that
we found about 94\% groups have significant elongation throughout the group and the surrounding
fields (GA$\ge$2), and a (negative) linear relation between GA and the fraction of $\eta_1$ galaxies. 
In the case when these objects are predominantly dEs, we conclude that these galaxies are tracing the
anisotropic large-scale accretion of matter onto groups. We also know that
the high dwarf-to-giant ratio observed 
in rich clusters suggests that cluster dE do not form
in groups that later merge to build clusters (Conselice 2005). 
Bright galaxies that follow the Kormendy relation (Kormendy 1977)
are indeed unlikely to have been formed by mergers of dwarf early-type systems (Evstigneeva et al. 2004). 
Likewise, our results indicate
that a high number of dEs exist in both the groups and the
flow of matter along the filamentary structure feeding these systems.

\section{Acknowledgments}
We thank the referee for useful suggestions.
We also thank A. Baddeley and B. Carvalho for the statistical tips.
A.L.B.R. thanks the support of CNPq, grants 201322/2007-2 and 471254/2008-8. 
P.A.A. Lopes  was supported by the Funda\c c\~ao de
Amparo \`a Pesquisa do Estado de S\~ao Paulo (FAPESP, processes 06/04955-1 and
07/04655-0).

\end{document}